# The geometry of chromatin

Subhash Kak
Chapman University and Oklahoma State University
Email: subhash.kak@okstate.edu

*Abstract*
We investigate the large-scale geometry of the DNA-protein complex of chromatin using a generalized optimality principle, which requires that not only should all sub-parts of a natural process be optimal but also the unfolding of higher recursive levels. It was shown previously that an information-theoretic geometry of the genetic code data, together with the principle of maximum entropy, explains the variation in the codon groupings that map into different amino acids and explain its underlying self-similar structure. Here we take that analysis forward and investigate the fundamental geometry underling physical and biological space as it gets reflected in aggregates associated with genomic DNA. The analysis is consistent with the measured fractal dimension of chromatin.

**Introduction**
Chromatin is a complex of tightly packed genomic DNA with proteins called histones that serve as spools around which DNA winds to create structural units called nucleosomes, which in turn are wrapped into 10-nanometer fibers that in the presence of cations fold into structures that are about 30 nanometer in diameter [1]. The nucleosome consists of about 147 bp of DNA wrapped 1.75 times around an octamer of core histone proteins.

Each histone-bound DNA molecule is a chromosome. It is dynamic, reorganizing itself during development and responding to environmental stimuli. Nuclear DNA does not appear in free linear strands. If each molecule of DNA were aligned end to end it would span nearly 2 meters, but these molecules fit into a cell nucleus that is approximately $6 \times 10^{-6}$ meters. Although it is packed tight, the spools do not get tangled and an array of pores in the nuclear membrane allows for the selective passage of certain proteins and nucleic acids into and out of the nucleus. Chromatin, also called the epigenome, not only compacts the genome into the nucleus, but also codes the mechanism controlling how the genome is read out from cell to cell.

As shown in Figure 1 (from [2]), 10-nm fibers are packaged in a fractal geometry so that the structural features of chromosomes are self-similar at many levels of magnification. This fractal behavior is a most significant characteristic of chromatin structure.

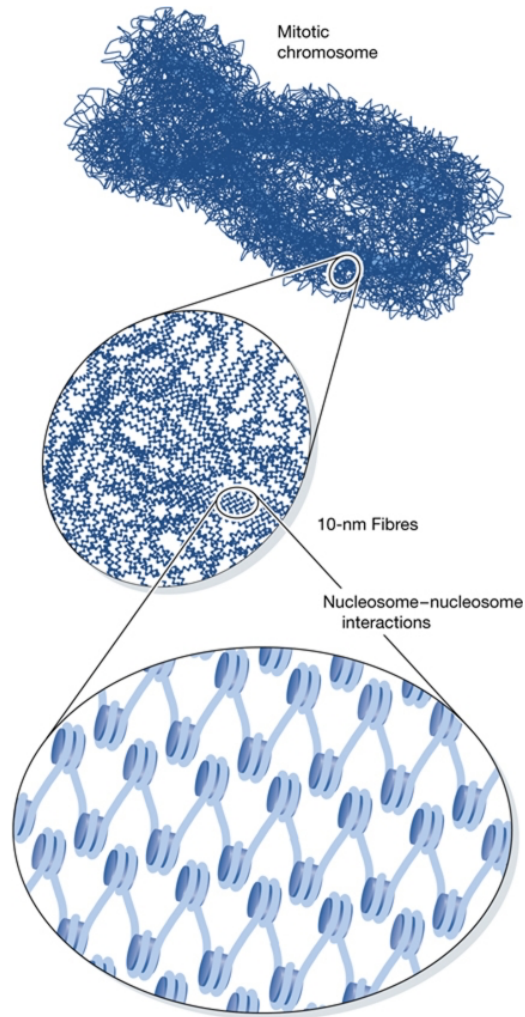

Figure 1. The fractal organization of chromatin (from [2])

Biological information may be seen both in the nature of the molecular structures and the higher-level symbolic representation of the codon sequences, and this mirrors the division in computers of hardware and software. Therefore, fractal structure should show up in the genetic code itself, as is suggested by the highly irregular mapping scheme between number of codons corresponding to different amino acids.

**Table 1** Number of codons to amino acid

| codons to amino acid | 1 | 2 | 3 | 4 | 5 | 6 |
|---|---|---|---|---|---|---|
| number of this type | 2 | 9 | 1 | 5 | 0 | 3 |

It was shown that the genetic code must be viewed not only on stereochemical, co-evolution, and error-correction considerations, but also on two additional factors of significance to natural systems, that of an information-theoretic geometry of the code data that leads to fractal behavior, and the principle of maximum entropy, and doing so explains the variation in the codon groupings [3] (see also [4][5]) and it can conceivably be used to examine evolutionary history.

This indicates that the perspective of information dimensionality be used not only for the code but also for larger aggregates of genetic materials and as required by the optimality principle this should work at all levels of recursion. In this paper, we use information theoretic optimality to examine the overarching geometrical constraints associated with the three-dimensional conformations of chromatin to determine its fractal dimension.

**Optimal coding of biological information**
In the mapping from RNA to amino acid sequence (Figure 2), there are 64 possible combinations of three-letter nucleotide sequences that can be made from the four nucleotides, which are the 64 codons; of these, 61 encode for amino acids, and 3 are signal to the ribosome to stop. The AUG codon, in addition to coding for the amino acid methionine, is found at the beginning of every messenger RNA and it indicates the start of a protein. Methionine and tryptophan are the only two amino acids that are coded by just a single codon (AUG and UGG, respectively).

There is a variation in the number of codons for each amino acid that ranges from 1 to 6 (Table 1) and it has been assumed that this variation is an artifact of the randomness of the mapping from the 61 codons into the 20 amino acids. But this departs radically from the anticipated 3 codons to an amino acid mapping that one would have expected on redundancy and symmetry considerations [6][7].

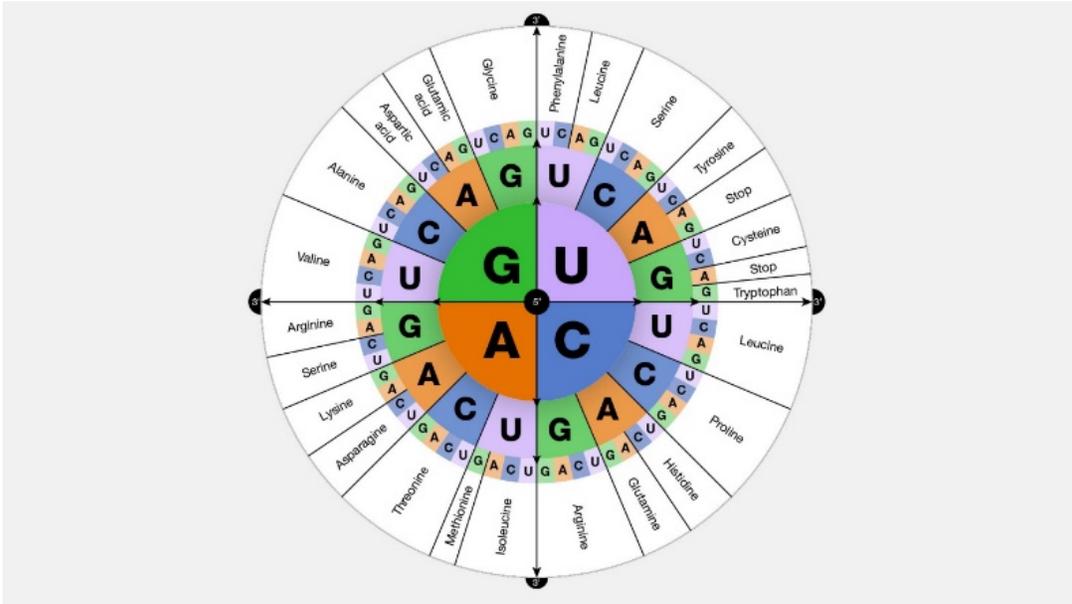

Figure 2. The genetic code where the codon sequence is read from center out.
https://www.genome.gov/sites/default/files/media/images/tg/Genetic-code.jpg

The idea of information theoretic geometry comes from a mathematical theorem according to which the optimal way to represent information is in $e$ ($\approx$ 2.718) classes [8], which has applications to a variety of fields that range from physics [9], cosmology [10], and large physical structures [11]. The noninteger dimensionality of this number represents a geometry that can be seen to be at the basis of fractal structures that are characterized by self-similarity and scale invariance [12].

According to the maximum entropy principle [13], which can be seen as a restatement of the second law of thermodynamics, probability is assigned to different arrangements associated with a system in a manner so that it requires maximal information to unscramble it, suggesting that nature's purpose may be viewed as the *concealment* of information.

The genetic code leads to the protein's three-dimensional conformation as in Figure 3. This quite parallels the organization of natural languages where the corresponding sequence is phonemes (or letters) to words to sentence, in which two codes are implicit: one related to the production of phonemes, and the second the grammar at the basis of the organization of words. The details of natural languages are determined through a process of co-evolution, and they contain many context-sensitive rules, and a simple coding table cannot express them. For a point of comparison, we know from the history of grammar that Pāṇini's grammar, the most successful effort ever in the complete description of a language, comprises of nearly 4,000 algebraic rules [14].

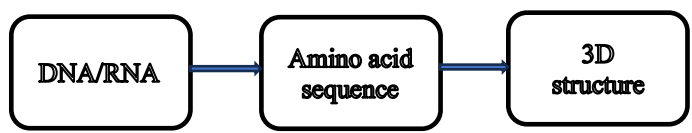

Figure 3. Three levels of the operation of the genetic code

The problem from the perspective of the genetic code is much simpler since the problem is clearly divided into two stages for in the words of Francis Crick, the genetic code is a small dictionary that relates the four letters of language of nucleic acids to the twenty-letter language of proteins [15]. Some denatured proteins fold to their native structures in only microseconds, on average, implying that there is a folding "mechanism," i.e., a particular set of events by which the protein finesses a broader conformational search, and this mechanism must work in the case of chromatin packing as well.

**Genetic code and protein folding**
A protein's amino acid sequence determines its characteristic three-dimensional conformation by which it becomes biologically functional [16]. But computational prediction of the structure is expensive for the number of different conformation states is astronomically large, even if they fold in subunits of 25 to 30 units [17][18]. Nevertheless, neural network and AI based techniques have become increasingly better at prediction of the structure [19][20]. Most recently, AlphaFold, a machine learning approach that incorporates physical and biological knowledge about protein structure and leverages multi-sequence alignments has demonstrated accuracy competitive with experimental systems [21]. But it doesn't provide the intermediate folding steps that may be crucial in finding answers to significant questions [22].

The diversity and specificity in function of proteins is made possible by the folding properties in which some lend rigidity to muscle cells or long thin neurons, and others bind to specific molecules and help them to new locations, and still others catalyze reactions that allow cells to divide and grow.

Synonymous codons also influence the function of proteins [23]. The genetic codes contain additional information [24][25] beyond amino acid sequences and the synonyms can have information on the rate of reaction [26].

The ordering of the amino acids determines the nature of the folding. Typically, two types of structures usually form in the secondary structure. Some regions coil up into formations called α-helices, which are formed by hydrogen bonding of the backbone to form a spiral shape, while other regions fold into β pleated sheets that are formed with the backbone bending over itself to form the hydrogen bonds. These two forms can interact to form more complex structures.

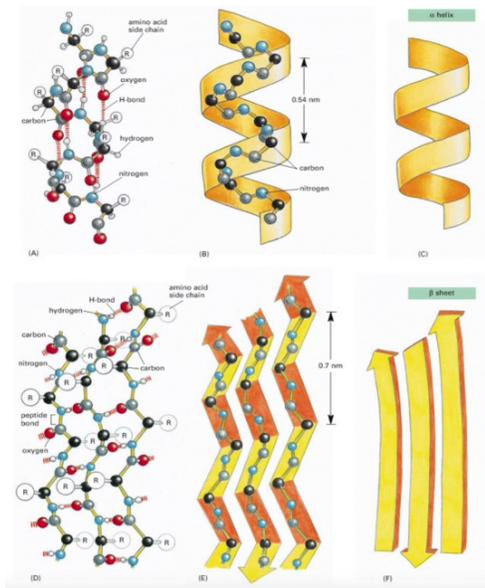

Figure 4. α-helices and β sheets B. Alberts et al. Molecular Biology of the Cell. Garland Science, 2002

The α-helices and β-sheets have a hydrophilic and a hydrophobic portion that helps in the forming tertiary structure in which the hydrophilic sides face the aqueous environment surrounding the protein and the hydrophobic sides face the hydrophobic core of the protein. From tertiary structure arise quaternary structure in some proteins, which involves the assembly or co-assembly of subunits that have already folded.

As the polypeptide chain is being synthesized by a ribosome, the linear chain begins to fold sometimes even during the translation of the polypeptide chain.

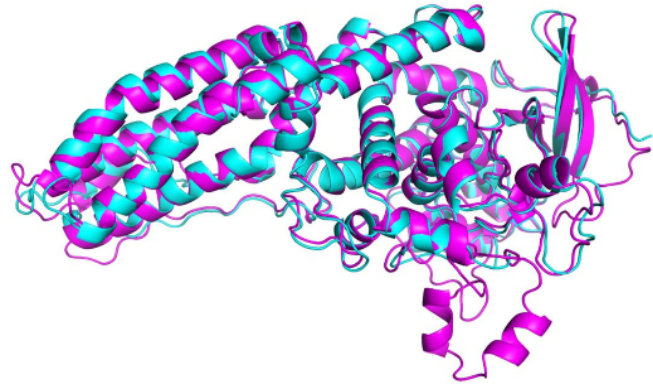

Figure 5. A folded protein; CASP / DEEPMIND
https://www.technologyreview.com/2020/11/30/1012712/deepmind-protein-folding-ai-solved-biology-science-drugs-disease/

Protein misfolding happens frequently inside of cells. Fortunately, there are several systems in place to refold or destroy aberrant protein formations. Molecular chaperones are a class of proteins that aid in the correct folding of other proteins in vivo. Another line of cell defense against misfolded proteins is called the proteasome. Proteasomes are protein complexes which degrade unneeded or damaged proteins by proteolysis, a chemical reaction that breaks peptide bonds.

**Recursive optimality of representation**
Although the *e*-dimensional informational space is optimal (as demonstrated in [3][8][9]), that doesn't tell us the intermediate states of the evolving system that starts off as a one-dimensional line structure.

Consider a recursive generalization of the optimality principle according to which the representation should be optimal at each recursive step.

Let there be a 3D system where we consider *k* cells in each direction in which we can see the folded stripes of the protein strands. For a maximal occupation of these total of $k^3$ cells, we must choose the appropriate size for if the size is too large, then all cells will be automatically filled, and if it is too small, then one would be considering size that is smaller than the characteristic geometry of the biological system.

If all cells are occupied, then the probability of each is $\frac{1}{k^3}$. The measure of information is the logarithm. In the binary world, we take it to the base 2, which is obvious when we see that for example 16 binary sequences can be represented by 4 bits counting from 0000, 0001, … 1111, that is $\ln_2 16$. If the base is $e$, and the total number of patterns is then the information in z patterns is $\ln z$. For the case being considered here, the number of cells is $k^3$, and, therefore, the information is $\ln k^3$.

Since the number of patterns is $k^3$, the efficiency of this system of information representation is

$$E(k) = \frac{\ln k^3}{k^3}$$

The system is optimal if $\frac{dE}{dk} = 0$. Differentiating, we obtain

$$\frac{3}{k^4} - \frac{3 \ln k}{k^4} = 0$$

It follows that $\ln k = 1$, or $k = e$.

Therefore, one would expect *e*-dimensionality in the aggregate representation of genetic code information, just as it is so when considering symbolic data. It is a pleasing result that physical and symbolic data should be characterized by the same constraint on optimality. It is the noninteger dimensionality of *e* that is the source of the self-similar fractal structure.

The self-similarity associated with noninteger dimensionality may be investigated in a variety of ways [27]. In earlier investigations, a fractal dimensional representation was found useful in the analysis of aggregate brain system as well [28][29]. One should note that computing the fractal dimension of aggregated microscopic structures requires that careful thought be given to the scales at which the measurements are made.

**Chromatin measurements**

Récamier et al. [30] have made careful measurements on chromatin fraction dimensionality and found the precise value of 2.7. Since the authors were not aware of the background theory of information-theoretic geometry, they didn't recognize the value as being a close approximation to *e*. This is what they conclude: "We computed the distribution of distances between every two points of the chromatin structure, and found that it followed a power law, leading to a precise measurement of the correlation fractal dimension of chromatin of 2.7. Moreover, we observed dynamic evolution of chromatin sub-regions compaction. as a result, the correlation fractal dimension of chromatin reported here can be interpreted as a dynamically maintained non-equilibrium state."

The chromatin is not packed tight, and the domains are not completely isolated from each other. The local density decreases as one travels to the periphery, and the packing domains may be seen as complex networks that are connected by chromatin fibers [31]. This aggregation structure may be seen as another coding scheme that is generated by the geometry of the genetic code.

It raises several questions such as determining the variability in the geometry within biological structures and implications of that for function. Thus, it has been suggested that the fractal characteristics of chromatin could be related to tumor pathology and pathophysiology [32][33].

Just as going from phonemes to words and sentences with meaning involves several other rules and mappings, one might ask if α-helices and β sheets for proteins and histones and nucleosomes for chromatin form intermediate structures that may be related at a fundamental level to the information geometry alphabet of biological information.

**Conclusions**

We used a generalized optimality principle, which requires that not only should all sub-parts of a natural process be optimal but also the evolution of its higher recursive levels, to investigate the large-scale geometry of the packing of chromatin. Specifically, we investigated the recursive geometry underling physical and biological space as it gets reflected in aggregations of matter. This framework yielded a result that is consistent with the measured fractal dimension of chromatin.

The research in this paper raises many new questions. It is necessary to investigate if helical systems and sheets represent intermediate forms in the passage beyond linearity. One would like to know the mathematical and logical bases to the emergence of these intermediate forms. The answer to these questions will further elucidate aspects of the packing of genomic DNA and histones.